\documentclass[twocolumn,showpacs,prb,aps, superscriptaddress]{revtex4}
\usepackage{amssymb}
\usepackage{amsmath}
\usepackage{graphicx}

\begin{document}

\title{Quantum impurity in the bulk of topological insulator}
\author{Hai-Feng L\"{u}}
\affiliation{Department of Physics, The University of Hong Kong, Pokfulam Road, Hong
Kong, China}
\affiliation{Department of Applied Physics, University of Electronic Science and
Technology of China, Chengdu 610054, China}
\author{Hai-Zhou Lu}
\affiliation{Department of Physics, The University of Hong Kong, Pokfulam Road, Hong
Kong, China}
\author{Shun-Qing Shen}
\affiliation{Department of Physics, The University of Hong Kong, Pokfulam Road, Hong
Kong, China}
\author{Tai-Kai Ng}
\affiliation{Department of Physics, Hong Kong University of Science and Technology, Clear
Water Bay Road, Hong Kong, China}
\date{\today}

\begin{abstract}
We investigate physical properties of an Anderson impurity embedded in the
bulk of a topological insulator. The slave-boson mean-field approximation is
used to account for the strong electron correlation at the impurity.
Different from the results of a quantum impurity on the surface of a
topological insulator, we find for the band-inverted case, that a Kondo resonant
peak and in-gap bound states can be produced simultaneously. However, only
one of them appears for the normal case. It is shown that the
mixed-valence regime is much broader in the band-inverted case, while it
shrinks to a very narrow regime in the normal case. Furthermore, a
self-screening of the Kondo effect may appear when the interaction between
the bound-state spin and impurity spin is taken into account.
\end{abstract}

\maketitle

\section{Introduction}

A topological insulator (TI) is 
insulating in the bulk, but hosts conducting edge or surface states near the
system boundary. It has attracted attention in the community of condensed
matter physics due to its potential application in spintronics and quantum
computation.\cite{Hasan10rmp,Qi11rmp,Moore10nature} A class of materials
such as Bi$_{2}$Se$_{3}$ and Bi$_{2}$Te$_{3}$ has been found to possess 
surface states which form a single Dirac cone.\cite%
{Zhang09natphys,Chen09science,Xia09natphys} Suppression of backscattering
inside the Dirac cone guarantees that the Dirac dispersion remains
essentially unperturbed for weak perturbation that preserves time-reversal
symmetry.\cite{Wu06prl,Xu06prb} So far, the effects of impurity
scattering on the surface of TIs have been investigated extensively. \cite%
{Liu09prl,Biswas10prb,Roushan09nature,Zhang09prl,Alpichshev10prl,
Alpichshev12prl,Schaffer12prb,Zitko10prb, Tran10prb,Liu09prb,Shan11prb,
Lu11njp,Wray10natphys,Feng10prb}
In the presence of classical spins which break time-reversal symmetry, it
was predicted that the impurity could open up a local gap and suppress the
local density of states.\cite{Liu09prl,Biswas10prb} The suppression of
backscattering around nonmagnetic impurities on surfaces with strong
spin-orbit coupling has been confirmed by scanning tunneling miscroscope
experiments.\cite{Roushan09nature,Zhang09prl,Alpichshev10prl,Alpichshev12prl}
However, strong nonmagnetic scattering, such as that from electrostatic potentials,
may disrupt the Dirac cone and create low-energy impurity resonances.\cite%
{Schaffer12prb} For a quantum impurity on the surface of a TI, the
Hamiltonian for the impurity can be mapped to the conventional pseudo-gap
Anderson model. The impurity is fully screened at low temperatures when the
Fermi level is located away from the Dirac point.\cite{Zitko10prb,Tran10prb}

Although there are theoretical and experimental studies on the
quasi-particle states around an impurity, most of them focus on the
impurity on the surface of the TI. Essentially, the topological nature of TIs
is determined by the electronic structure of the bulk bands instead of the
surface states. On the other hand, the TI samples available nowadays are
always poorly insulating in the bulk, owing to a large amount of vacancies
and defects.\cite{Checkelsky09prl,Eto10prb,Ren10prb,Qu10science,Peng10natmat}
For these reasons, it is important to study how the quasi-particle states
are affected when vacancies or impurities are localized in the bulk of the
system. It has been shown that classical spins \cite{Liu09prb} or vacancies%
\cite{Shan11prb,Lu11njp} localized in the bulk of TIs could result in the
coexistence of in-gap bound states and boundary states. For a quantum
impurity, the quantum fluctuations of its internal degree of freedom play an
important role, making it significantly differ from classical impurities.%
\cite{Balatsky06rmp} However, it remains unknown how a quantum impurity in
the bulk of a TI affects the electronic states.

The study of quantum impurity in TIs is also related to
the problems of impurities in unconventional density waves,\cite%
{Balatsky06rmp,Withoff90prl,Buxton98prb,Dora05prb} gapped systems,\cite%
{Ogura93jpsj,Galpin08prb} or spin-orbit-coupled systems.\cite%
{Zarea12prl,Zitko11prb,Eriksson12prb} For instance, in a gapped system,
the Kondo effect breaks down when the energy gap exceeds a critical
value.\cite{Ogura93jpsj} In the presence of Rashba spin-orbit
interaction, a parity-breaking Dzyaloshinsky-Moriya term could be
induced, resulting in a possible change of the Kondo
temperature.\cite{Zarea12prl,Zitko11prb,Eriksson12prb} Since there are strong
spin-orbit couplings and unconventional gaps in the TIs, it becomes
interesting to investigate the differences between the Kondo effects
in conventional insulators and TIs.

In this paper, we investigate the effects of a quantum impurity embedded in
a TI with the help of the slave-boson mean-field approach. We show that
in-gap bound states and Kondo effect could coexist in TIs, while only one
of them appears for conventional insulators. If the bound states are
singly occupied, the Kondo resonance could be screened by the exchange
interaction between the impurity spin and the spin of the impurity-induced
bound states, leading to a self-screened Kondo effect. The paper is
organized as follows. In Sec. II we introduce a model Hamiltonian of an
impurity in a TI and the slave-boson mean-field approach. In Sec. III we
discuss the Kondo effect and the formation of the in-gap bound states in
both band-inverted and normal cases. In Sec. IV, we show the self-screening
of the Kondo effect. Finally, a summary is presented in Sec. V.

\section{Model Hamiltonian and slave boson approach}

\subsection{Model}

The effective model to describe the bulk states of the TI with an impurity is
written as
\begin{equation}
H=H_{0}+H_{d}+H_{t}.
\end{equation}%
The part for the bulk electrons of TI is given by the modified Dirac model%
\cite{Lu10prb, Shan10njp, Shen11spin}
\begin{equation*}
H_{0}=\Psi _{k}^{\dagger }[\hbar v_{F}\vec{k}\cdot \vec{\alpha}%
+(mv_{F}^{2}-B\hbar ^{2}k^{2})\beta ]\Psi _{k}
\end{equation*}%
with $\alpha _{i}=\sigma _{x}\otimes \sigma _{i}$ and $\beta =\sigma
_{z}\otimes \sigma _{0}$, where $\sigma _{i}$ $(i=x,y,z)$ are the Pauli
matrices and $\sigma _{0}$ is the $2\times 2$ unit matrix. $k_{i}=-i\partial
_{i}$ ($i=x,y,z$) is the momentum operator, $%
k^{2}=k_{x}^{2}+k_{y}^{2}+k_{z}^{2}$, and $v_{F}$ and $m$ have the dimensions of
speed and mass, respectively. Different from the surface Hamiltonian, a
quadratic correction in momentum $-B\hbar ^{2}k^{2}$ and a gap term $%
mv_{F}^{2}$ are introduced in the bulk Hamiltonian. The sign of $mB$
determines whether the system is topologically trivial or not: it is
nontrivial for $mB>0$ (i.e., the band-inverted case), and trivial for $mB<0$
(i.e., the normal case).\cite{Lu10prb,Shan10njp,Shen11spin} The energy
spectra have a finite energy gap. Here the basis vectors are chosen as
\begin{equation*}
\Psi _{k}^{\dagger }=\left( a_{k\uparrow }^{\dagger },a_{k\downarrow
}^{\dagger },b_{k\uparrow }^{\dagger },b_{k\downarrow }^{\dagger }\right) ,
\end{equation*}%
where $a_{k\sigma }^{\dagger }$ and $b_{k\sigma }^{\dagger }$ are creation
operators of electrons with spin $\sigma $ on two different orbits. In this
representation, we can rewrite the total Hamiltonian in the second-quantized
form
\begin{eqnarray}
H_{0} &=&\sum_{k\sigma }(mv_{F}^{2}-B\hbar ^{2}k^{2})(a_{k\sigma }^{\dagger
}a_{k\sigma }-b_{k\sigma }^{\dagger }b_{k\sigma })  \notag \\
&&+\hbar v_{F}\sum_{k}[k_{z}(a_{k\uparrow }^{\dagger }b_{k\uparrow
}-a_{k\downarrow }^{\dagger }b_{k\downarrow })  \notag \\
&&+(k_{x}-ik_{y})(a_{k\uparrow }^{\dagger }b_{k\downarrow }+b_{k\uparrow
}^{\dagger }a_{k\downarrow })+\mathrm{H.c.}].
\end{eqnarray}%
The Hamiltonian that describes the Anderson impurity is
\begin{equation}\label{Hd}
H_{d}=\epsilon _{d}(c_{d\uparrow }^{\dagger }c_{d\uparrow }+c_{d\downarrow
}^{\dagger }c_{d\downarrow })+Uc_{d\uparrow }^{\dagger }c_{d\uparrow
}c_{d\downarrow }^{\dagger }c_{d\downarrow },
\end{equation}%
where $\epsilon _{d}$ is the impurity energy level and $U$ is the on-site
Coulomb interaction. The coupling Hamiltonian between the impurity and
electrons in TI has the form
\begin{equation}\label{Ht}
H_{t}=\sum_{k\sigma }(V_{ak}a_{k\sigma }^{\dagger }c_{d\sigma
}+V_{bk}b_{k\sigma }^{\dagger }c_{d\sigma }+\mathrm{H.c.}),
\end{equation}%
where $V_{a(b)k}$ represents the overlap or hybridization matrix
element between the magnetic impurity and conduction electrons in
two bands.

\subsection{\label{sec:sbmf}Slave-boson mean field}

Here we consider a strong on-site Coulomb interaction on the impurity, i.e.,
$U\rightarrow \infty $. In this limit, no double occupancy on the impurity
is allowed. We introduce the auxiliary fields
\begin{equation*}
c_{d\sigma }^{\dagger }=d_{\sigma }^{\dagger }b,c_{d\sigma }=b^{\dagger
}d_{\sigma },
\end{equation*}%
where the boson operator $b^{\dagger }$ creates an empty state and the
fermion operator $d_{\sigma }^{\dagger }$ creates a singly occupied state on
the impurity. These two fields obey the local constraint $b^{\dagger
}b+\sum_{\sigma }d^{\dagger }d=1$.\cite{Hewson93book} In the mean-field
approximation, both the annihilation and creation boson operators $b$ and $%
b^{\dagger }$ are replaced by a complex number $b_{0}$ and its complex
conjugate $b_{0}^{\ast },$ and the local constraint is realized by
introducing a Lagrangian multiplier $\lambda _{0}$. Substituting the 
auxiliary fields in the
original Hamiltonians (\ref{Hd}) and (\ref{Ht}), one can get
\begin{equation*}
H_{d}=\tilde{\epsilon}_{d}(d_{\uparrow }^{\dagger }d_{\uparrow
}+d_{\downarrow }^{\dagger }d_{\downarrow })+\lambda _{0}(b_{0}^{2}-1)
\end{equation*}%
and
\begin{equation*}
H_{t}=\sum_{k\sigma }(\tilde{V}_{ak}a_{k\sigma }^{\dagger }d_{\sigma }+%
\tilde{V}_{bk}b_{k\sigma }^{\dagger }d_{\sigma }+\mathrm{H.c.})
\end{equation*}%
with the renormalized parameters $\tilde{\epsilon}_{d}=\epsilon _{d}+\lambda
_{0}$ and $\tilde{V}_{a(b)k}=b_{0}V_{a(b)k}$. The slave-boson mean-field
approximation was first introduced to describe the low-energy physics of
the conventional Anderson impurity model in the mixed-valence regime.\cite%
{Coleman84prb} This method may produce the low-energy physics in
unconventional density-waves (e.g., $d$-wave superconductors,\cite{Zhu00prb}
graphene electron systems,\cite{Zhu11prb} etc.).

\subsection{Green's functions}

Utilizing the method of the equation of motion for the impurity electron,
one finally obtains the retarded Green's function of the impurity electron,
\begin{equation}
\langle \langle d_{\sigma }|d_{\sigma }^{\dagger }\rangle \rangle =\frac{%
\omega -\tilde{\epsilon}_{d}-\Sigma _{0}(\omega )+\sigma \Sigma _{z}(\omega )%
}{[\omega -\tilde{\epsilon}_{d}-\Sigma _{0}(\omega )]^{2}-\sum_{i}\left[
\Sigma _{i}(\omega )\right] ^{2}},
\end{equation}%
where the self-energy functions are defined as
\begin{eqnarray}
\Sigma _{0}(\omega ) &=&\sum_{k}\frac{(\omega +A_{k})\tilde{V}%
_{ak}^{2}+(\omega -A_{k})\tilde{V}_{bk}^{2}}{\omega ^{2}-\hbar
^{2}v_{F}^{2}k^{2}-A_{k}^{2}},  \notag \\
\Sigma _{i}(\omega ) &=&\sum_{k}\frac{2\hbar v_{F}k_{i}\tilde{V}_{ak}\tilde{V%
}_{bk}}{\omega ^{2}-\hbar ^{2}v_{F}^{2}k^{2}-A_{k}^{2}},
\end{eqnarray}%
($i=x,y,z$), and $A_{k}=mv_{F}^{2}-B\hbar ^{2}k^{2}$ .

It is assumed that the hybridization strength $V_{a(b)k}$ does not depend on
momentum, $V_{a(b)k}=V_{a(b)}$. In the absence of an external magnetic field,
the interaction self-energy $\Sigma_i(\omega)=0$ and the electronic Green's
function can be simplified as
\begin{eqnarray}
\langle\langle d_{\sigma}|d_\sigma^{\dagger }\rangle\rangle=\frac{1} {\omega-%
\tilde{\epsilon}_d-\Sigma_0(\omega)},
\end{eqnarray}%
which has the same form as those of the conventional Anderson impurity
model. Additionally, we consider the case that the impurity is symmetrically
coupled with two orbits, i.e., $V_0=V_a=V_b$, and the self-energy $%
\Sigma_0(\omega)$ in $d$-dimensions ($d=2, 3$) is thus simplified as
\begin{eqnarray}
\Sigma_0(\omega)&=&\omega\sum_k\frac{2\tilde{V}_0^2}{\omega^2
-\hbar^2v_F^2k^2-A_k^2}  \notag \\
&=&\frac{d\omega \tilde{V}_0^2N_0}{k_F^d}\int_0^\infty dk\frac{k^{d-1}} {%
\omega^2-\hbar^2v_F^2k^2-A_k^2},
\end{eqnarray}%
where $N_0$ is the number of lattice sites and $k_F$ is the Fermi wave
vector for the bulk of a TI.

\subsection{Normal and band-inverted regimes}

From the poles of self-energy function $\Sigma _{0}(\omega )$, one obtains
the energy spectrum of the bulk,
\begin{equation*}
\xi (k)=\pm \sqrt{\hbar ^{2}v_{F}^{2}k^{2}+(mv_{F}^{2}-B\hbar ^{2}k^{2})^{2}}%
.
\end{equation*}%
The energy spectrum shows a complex dependence on the energy gap and
momentum. Such energy spectra can be mapped onto many important cases of
impurity problems. Near the band edges, the model reduces to the Anderson
problem in normal insulators or semiconductors.\cite{Ogura93jpsj,Galpin08prb}
The quantum impurity model in graphene or surface states of the TIs can be
recovered by setting $m=0$ and $B=0$.\cite%
{Zitko10prb,Tran10prb,Feng10prb,Zhu11prb}

\begin{figure}[htbp]
\centering
\includegraphics[width=0.5\textwidth]{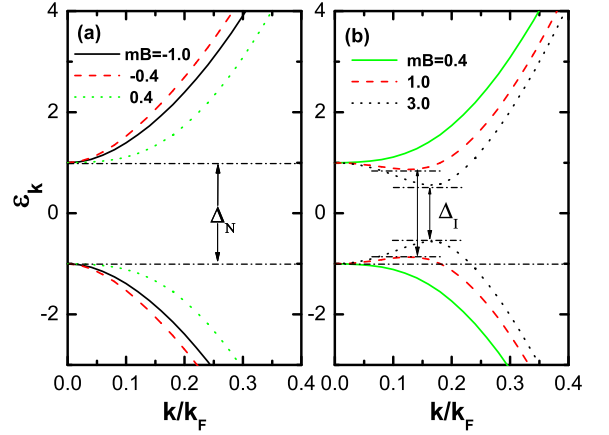}
\caption{The energy spectrum $\epsilon_k$ for different $mB$ values: (a) $mB=-1.0$, $-0.4$, $0.4$;
(b) $mB=0.4$, $1.0$, $3.0$. Here $mv_F^2$
is taken as the energy unit. For $mB<1/2$, the energy gap is located at $k=0$ and equals
 $\Delta _{N}\equiv 2|m|v_{F}^{2}$. For $mB>1/2$, the energy gap is located at a
finite $k$ and equals
 $\Delta _{I}\equiv \frac{\sqrt{4mB-1}}{|B|}v_{F}^{2}$.}
\label{fig-1}
\end{figure}

In Fig. 1 we present the energy spectra as a function of $k$ for
different values of $mB$. It can be seen from the energy spectrum
that, for $mB<1/2$, the band edges are located at $k=0$ and the
energy gap is
\begin{equation*}
\Delta _{N}\equiv 2|m|v_{F}^{2}
\end{equation*}%
as in normal insulators. However, for the case of $mB>1/2$, the band edges
appear at a finite $k$ with a gap
\begin{equation*}
\Delta _{I}\equiv \frac{\sqrt{4mB-1}}{|B|}v_{F}^{2}.
\end{equation*}%
For convenience, one defines
\begin{equation}
g_{0}(\epsilon )=\frac{dN_{0}\tilde{V}_{0}^{2}\epsilon \left[ \sqrt{\epsilon
^{2}-\Delta _{I}^{2}/4}-\frac{1-2mB}{2|B|}v_{F}^{2}\right] ^{\frac{d-2}{2}}}{%
4(|B|\hbar ^{2}k_{F}^{2})^{\frac{d}{2}}\sqrt{\epsilon ^{2}-\Delta _{I}^{2}/4}%
}
\end{equation}%
when $\epsilon >\Delta _{N}/2$, and
\begin{eqnarray}
g_{1}(\epsilon ) &=&\frac{dN_{0}\tilde{V}_{0}^{2}\epsilon \left[ \sqrt{%
\epsilon ^{2}-\Delta _{I}^{2}/4}+\frac{2mB-1}{2|B|}v_{F}^{2}\right] ^{\frac{%
d-2}{2}}}{4(|B|\hbar ^{2}k_{F}^{2})^{\frac{d}{2}}\sqrt{\epsilon ^{2}-\Delta
_{I}^{2}/4}}  \notag \\
&&
\end{eqnarray}%
when both $\Delta _{I}/2<\epsilon <\Delta _{N}/2$ and $mB>1/2$ are satisfied.

For the case of $mB<1/2$, one gets
\begin{equation}
\Sigma _{0}(\omega )=\int_{\frac{\Delta _{N}}{2}}^{\infty }d\xi g_{0}(\xi )%
\left[ \frac{1}{\omega -\xi }+\frac{1}{\omega +\xi }\right] .
\end{equation}%
After analytical continuation, $\Sigma _{0}(\omega )=\mathrm{Re}\Sigma
_{0}(\omega )+\mathrm{Im}\Sigma _{0}(\omega )$ with
\begin{equation}
\mathrm{Re}\Sigma _{0}(\omega )=\mathrm{P}\int_{\frac{\Delta _{N}}{2}%
}^{\infty }d\xi g_{0}(\xi )\left[ \frac{1}{\omega -\xi }+\frac{1}{\omega
+\xi }\right]
\end{equation}%
and
\begin{equation}
\mathrm{Im}\Sigma _{0}(\omega )=-g_{0}(\omega )\Theta (|\omega |-\frac{%
\Delta _{N}}{2})
\end{equation}%
with $\mathrm{P}$ denoting the principal value.

For the case of $mB>1/2$, the real and imaginary parts of the self-energy are
given by
\begin{eqnarray}
\mathrm{Re}\Sigma _{0}(\omega ) &=&\mathrm{P}\left[ \int_{\frac{\Delta _{I}}{%
2}}^{\frac{\Delta _{N}}{2}}d\xi g_{2}(\xi )\right.   \notag \\
&+&\left. \int_{\frac{\Delta _{N}}{2}}^{\infty }d\xi g_{0}(\xi )\right]
\left( \frac{1}{\omega -\xi }+\frac{1}{\omega +\xi }\right)
\end{eqnarray}%
and
\begin{eqnarray}
\mathrm{Im}\Sigma _{0}(\omega ) &=&-\left[ g_{2}(\omega )\Theta (|\omega |-%
\frac{\Delta _{I}}{2})\Theta (\frac{\Delta _{N}}{2}-|\omega |)\right.
\notag \\
&&+\left. g_{0}(\omega )\Theta (|\omega |-\frac{\Delta _{N}}{2})\right] ,
\end{eqnarray}%
where $g_{2}(\omega )=g_{0}(\omega )+g_{1}(\omega )$. Because $\mathrm{Im}%
\Sigma _{0}$ is proportional to the density of states of the bulk electrons,
the density of states is discontinuous at the point $\Delta _{N}/2$. For a
relatively large $B$ and small $m$, the energy gap reduces to a value much
smaller than $\Delta _{N}$.

The free energy $F$ of the system is given by the partition function
\begin{eqnarray*}
F=-\frac{1}{\beta}\mathrm{ln}Z=-\frac{1}{\beta}\int_{-\infty}^
{\infty}\mathrm{ln}(1+e^{-\beta(\omega-\mu)})\bar{\rho}(\omega)d\omega,
\end{eqnarray*}%
where $\bar{\rho}(\omega)=\Sigma_k\delta(\omega-\epsilon_k)$,
$\epsilon_k$ are the one-electron energies of the system, $\mu$ is the
chemical potential of the TI, and $\beta=1/k_BT$ is the system
temperature. $\bar{\rho}(\omega)$ can be calculated from the poles
of the one-electron Green's function and is given by
\begin{eqnarray*}
\bar{\rho}(\omega)&=&\frac{\mathrm{Im}}{\pi}\sum_k\frac{4
\omega}{\omega^2-\hbar^2v_F^2k^2-A_k^2} +2\frac{\mathrm{Im}}{\pi}
\frac{\partial}{\partial \omega}
\mathrm{ln}
\langle\langle d_{\sigma}|d_{\sigma}^{\dagger }\rangle\rangle.
\end{eqnarray*}%

Minimizing the free energy of the system with respect to $\lambda _{0}$ and $%
b_{0}$, one obtains a set of self-consistent equations
\begin{eqnarray}
2\int_{-\infty }^{\infty }d\omega f(\omega )\rho _{d}(\omega )+b_{0}^{2}-1
&=&0,  \notag \\
2\int_{-\infty }^{\infty }d\omega f(\omega )(\omega -\tilde{\epsilon}%
_{d})\rho _{d}(\omega )+\lambda _{0}b_{0}^{2} &=&0,
\end{eqnarray}%
where $f(\omega )=[1+e^{(\omega -\mu )/k_{B}T}]^{-1}$ is the Fermi
distribution function and the density of states $\rho _{d}(\omega )$
of the impurity is given by
\begin{equation*}
\rho _{d}(\omega )=-\frac{1}{\pi }\frac{\mathrm{Im}\Sigma _{0}(\omega )}{%
[\omega -\tilde{\epsilon}_{d}-\mathrm{Re}\Sigma _{0}(\omega )]^{2}+\mathrm{Im%
}\Sigma _{0}(\omega )^{2}}.
\end{equation*}

In the calculation, it is limited to zero temperature, i.e., $k_{B}T=0$. $%
\Delta _{N}/2$ is taken as the energy unit and $\Gamma _{0}=\pi \rho
_{0}V_{0}^{2}$ represents the hybridization strength between the impurity
and the bulk electrons, where $\rho _{0}=N_{0}/2D$ is the density of states
per spin at the chemical potential and $D=\sqrt{(\hbar
v_{F}k_{F})^{2}+(mv_{F}^{2}-B\hbar ^{2}k_{F}^{2})^{2}}$ is a cut-off of the
band width. In the following, we show the results of quantum impurity in
three-dimensional TIs for $\Gamma _{0}=0.5$ and $D=30.0$.

\section{In-gap bound states and the Kondo effect}

\subsection{Self-energy}

First, we discuss the self-energy $\Sigma _{0}(\omega )$, which depends on
the energy spectrum of the bulk bands. In the bulk of the TI, the strong
spin-orbit coupling couples the conduction and valence bands, leading to the
non-parabolic energy spectrum. The self-energy $\Sigma _{0}(\omega )$
reveals not only the poles of the Green's function, but also the
inhomogeneous density of states due to the impurity-induced states.

\begin{widetext}

\begin{figure}[htbp]
\centering
\includegraphics[width=1.2\textwidth]{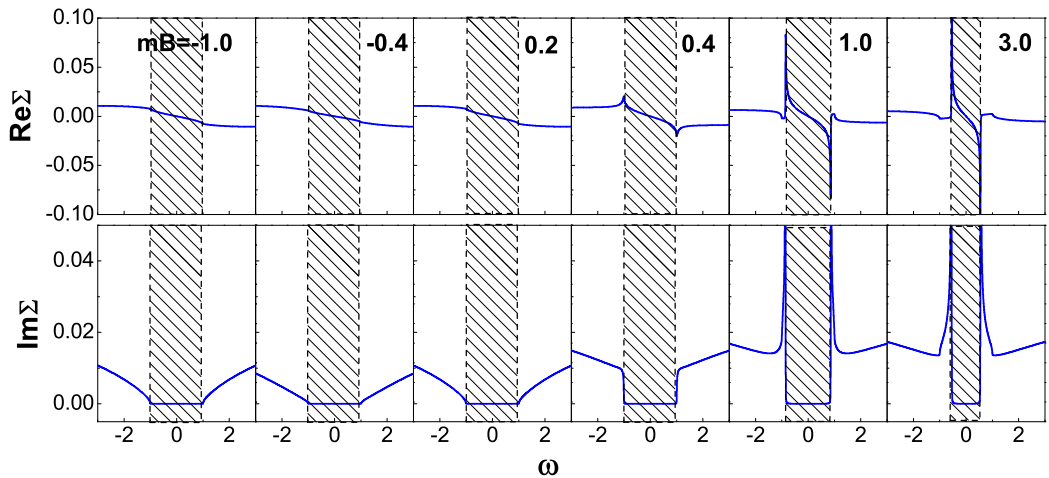}
\caption{The real and imaginary parts of the self-energy $\Sigma_0(\omega)$
for different $mB$ values. The shaded part represents the region of energy gap.}
\label{fig-2}
\end{figure}

\end{widetext}

Figure \ref{fig-2} shows the real and imaginary parts of $\Sigma
_{0}(\omega )$ versus $\omega $ for different values of $mB$.
Previous studies based on the Chern number and Z$_{2}$ invariant
indicate that the band-inverted case with $mB>0$ is topologically
nontrivial, while the normal case with $mB<0$ is topologically
trivial.\cite{Lu10prb,Shan10njp,Shen11spin} It is shown that in Fig.
\ref{fig-2} both the real and imaginary parts of $\Sigma _{0}(\omega
)$ are quite different for opposite signs of $mB$. The real part $\mathrm{Re}%
\Sigma _{0}(\omega )$ determines the level positions of both the Kondo
resonance and the bound states, according to the solutions to the equation
\begin{equation}\label{resonances}
\omega =\tilde{\epsilon}_{d}+\mathrm{Re}\Sigma _{0}(\omega ).
\end{equation}%
For $mB>1/2$, $\mathrm{Re}\Sigma _{0}(\omega )$ diverges rapidly near the
edges of the gap, thus there always exist bound states within the gap.
When the impurity level $|\epsilon _{d}|$ is much larger than $\Delta _{I}/2$%
, the bound states are very close to the bottom of the conduction bands for
$\epsilon _{d}\gg \Delta _{I}/2$ or the top of the valence band for $%
\epsilon _{d}\ll -\Delta _{I}/2$, which is similar to the case of $s$-wave
superconductors.\cite{Machida72ptp} More importantly, Eq. (\ref{resonances})
has an extra solution in the band region, corresponding to the coexisting
Kondo resonance. For $mB<0$, there is only one solution to Eq. (\ref%
{resonances}), so the Kondo effect and the bound states do not appear
simultaneously.

The imaginary part $\mathrm{Im}\Sigma_0(\omega)$ is proportional to the
density of states of bulk electrons. For $mB>1/2$, the energy gap is $%
\Delta_I$, thus $\mathrm{Im}\Sigma_0(\omega)=0$ in the region of $\omega\in[%
-\Delta_I/2, \Delta_I/2]$. At the band edges $\pm \Delta_I/2 $, $\mathrm{Im}%
\Sigma_0(\omega)$ shows divergences. In the region $|\omega|\in [\Delta_I/2,
\Delta_N/2]$, the divergences drops rapidly with the increasing $|\omega|$.
For $|\omega|>\Delta_N/2$, $\mathrm{Im}\Sigma_0(\omega)$ begins to increase
as a function of $\sqrt{\omega}$. For $0<mB<1/2$, the divergences at the
band edges $\pm \Delta_N/2$ disappear gradually with decreasing $mB$. For $%
mB<0$, $\mathrm{Im}\Sigma_0(\omega)$ at the band edges is always zero and
increases as a function of $\sqrt{\omega}$ for $|\omega|>\Delta_N/2$.

\begin{widetext}

\begin{figure}[htbp]
\centering
\includegraphics[width=0.9\textwidth]{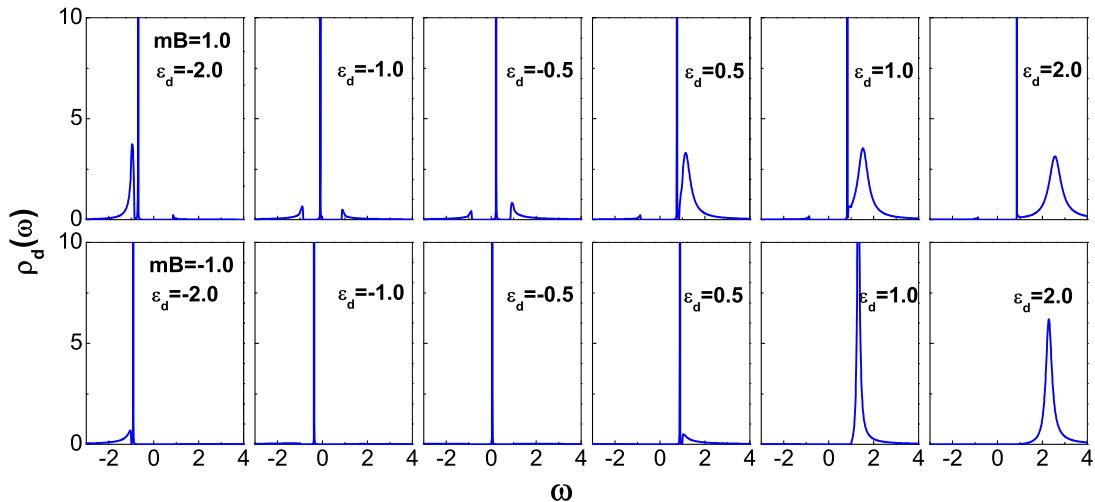}
\caption{ The density of states of the impurity electron is shown for
different  impurity levels $\epsilon_d=-2.0$, $-1.0$, $-0.5$, $0.5$, $1.0$, and $2.0$ (from left to right) in (a)
topological nontrivial case $mB=1.0$ and (b)
topological trivial case $mB=-1.0$, where the chemical potential
$\mu=0.0$ lies in the gap.}
\label{fig-3}
\end{figure}

\begin{figure}[htbp]
\centering
\includegraphics[width=0.9\textwidth]{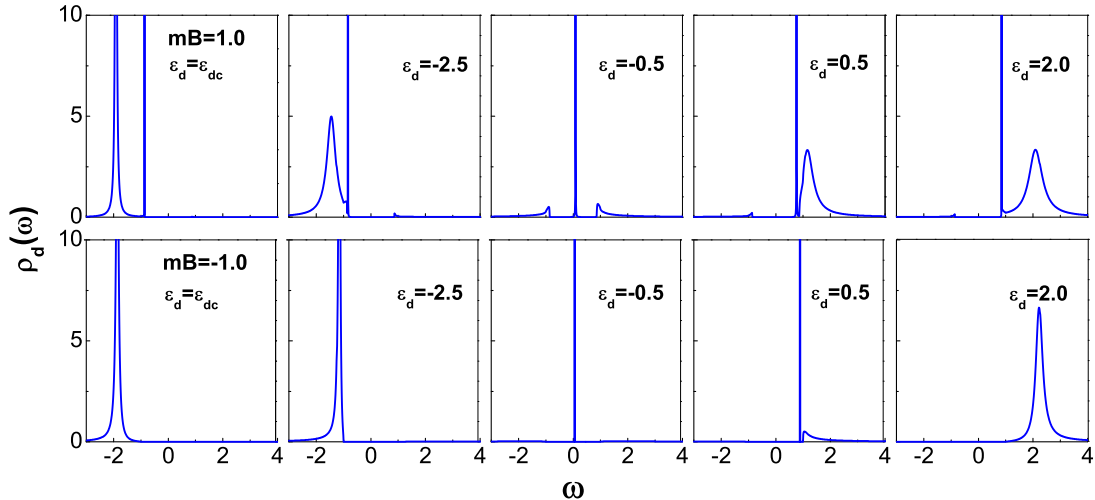}
\caption{ The density of states of the impurity electron is shown for
different  impurity levels $\epsilon_d=\epsilon_{dc}$ ($\epsilon_{dc}$ is the critical point at which $\tilde{\epsilon}_d=\mu$), $-2.5$, $-0.5$,
 $0.5$, and $2.0$ (from left to right) in (a) topological nontrivial case
$mB=1.0$ and (b) topological trivial case $mB=-1.0$, where the chemical potential $\mu=-2.0$ lies in the valence band.}
\label{fig-4}
\end{figure}

\end{widetext}

\begin{figure}[htbp]
\centering \includegraphics[width=0.5\textwidth]{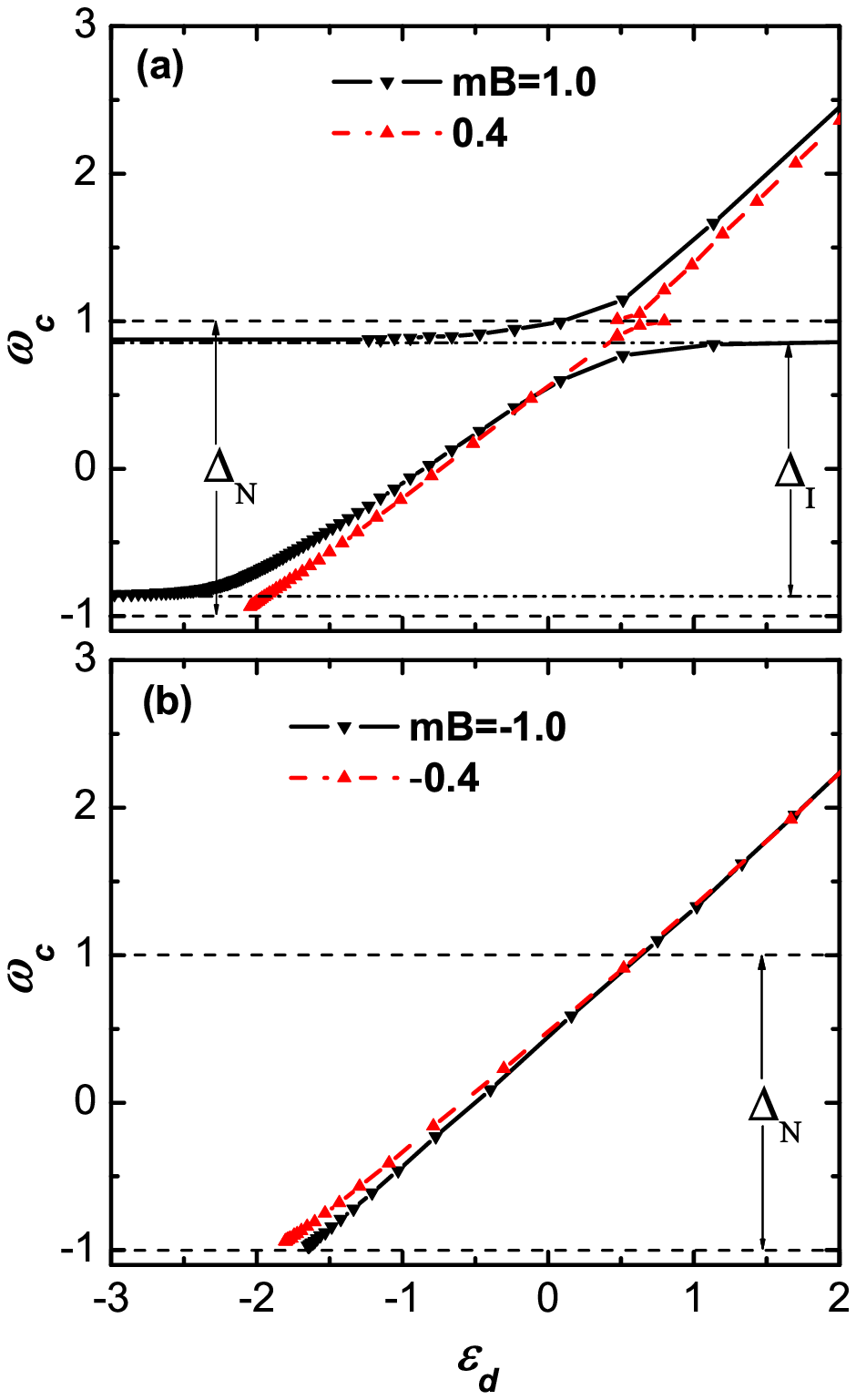}
\caption{The positions of the bound states and low-energy Kondo resonance
peak as functions of impurity level $\protect\epsilon_d$ in (a) the
topological nontrivial case $mB=1.0, 0.4$ and (b) the topological trivial case $mB=-1.0, -0.4$.
The chemical potential is taken as $\protect\mu=0.0$.}
\label{fig-5}
\end{figure}

\subsection{Density of states of the impurity}

From the self-energy, the Kondo resonance and the bound states have
been discussed qualitatively. The density of states $\rho
_{d}(\omega )$ of the impurity for the cases that the chemical
potential $\mu $ lies in the gap and in the valence bands are
presented in Figs. \ref{fig-3} and \ref{fig-4}, respectively. It is
demonstrated that in both cases the Kondo resonance and bound states
coexist for $mB>0$, while only one of them appears for $mB<0$.
For $mB>0$, the bound states are very close to the band edges when $\tilde{%
\epsilon}_{d}$ is far away from the energy gap. In the previous studies,\cite%
{Dora05prb,Ogura93jpsj,Galpin08prb} it has been argued that when the energy
gap exceeded an critical value ($\Delta /T_{K}=2.0$ predicted by slave-boson
mean-field theory\cite{Ogura93jpsj} and $T_{K}$ is the Kondo temperature),
the Kondo effect no longer  appears in the insulators with an
energy-independent density of states. However, for a complex dispersion
relation, the density of states is strongly energy-resolved and quite
different near the band edges for the band-inverted and normal cases. In the
band-inverted case, $\rho _{d}(\pm \Delta _{I}/2)$ equals zero exactly, due
to the divergence of $\mathrm{Im}\Sigma _{0}(\omega )$ near the band edges.
For $|\omega |\gg \Delta _{I}$, $\rho _{d}(\omega )$ also approaches
zero. Therefore, no matter how $\mu $ lies in the gap or in the bands,
there are always two low-energy resonance peaks lying near the band edges
for $mB>1/2$, as shown in the upper panels of Figs. \ref{fig-3} and \ref%
{fig-4}. When the chemical potential $\mu $ lies in the valence
band, the low-energy resonance peak becomes narrow and its position
moves to $\mu $ with decreasing $\epsilon _{d}$. Near the critical
point $\epsilon _{dc}$, the resonance peak becomes very sharp and
close to the chemical potential.

In Fig. 5 we present the positions of the in-gap bound states and
low-energy Kondo resonance as functions of the impurity level
$\epsilon _{d}$ in the normal and band-inverted cases. The bound
states show quite different behaviors in the two cases. For the
normal case, the position of the bound states is almost linear in
$\epsilon _{d}$. The in-gap bound states start from the point
$\epsilon _{dc}=-\Delta _{N}/2$, corresponding to the top of the
valence bands. At the bottom of the conduction bands, the bound
states enter the conduction bands continuously, and then the
low-energy Kondo resonance peak appears. However, for the
band-inverted case, the bound states in the gap and the Kondo
resonance can form simultaneously for the impurity level $\epsilon
_{d}>\epsilon _{dc}$. In this case, the positions of these two
states do not connect at any point. For an $\epsilon _{d}$ far away
from the chemical potential, the in-gap bound states are very close
to the band edges. Thus they are vulnerable to small perturbation or
thermal fluctuation, and may merge into the conduction bands easily.

Above we demonstrate that the presence of the in-gap bound states is
determined by the topological nature of the TI. For real TI
materials such as Bi$_{2}$Se$_{3}$ and Bi$_{2}$Te$_{3}$, the system
parameters from first principles
calculations\cite{Zhang09natphys} are $(mv_F^{2}, \hbar v_F, B)= (0.28$eV, $3.2$eV\AA, $%
33$eV\AA $^{2})$ and $(0.30$eV, $2.9$eV\AA, $%
57$eV\AA $^{2})$, respectively. Correspondingly, $mB\sim0.9$ for
Bi$_{2}$Se$_{3}$ and $2.0$ for Bi$_{2}$Te$_{3}$. Therefore, it is
expected that the coexistence of in-gap bound state and Kondo effect
could be observed in these two materials.

\subsection{Broadened mixed-valence regime for band-inverted case}

\begin{figure}[htbp]
\centering \includegraphics[width=0.5\textwidth]{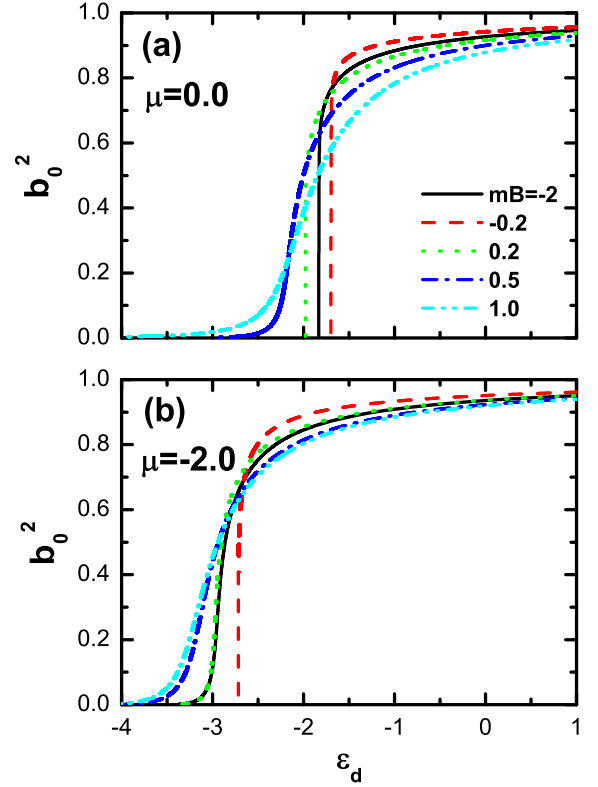}
\caption{The dependence of the order parameter $b_0^2$ on the impurity
energy level $\protect\epsilon_d$ for (a) the chemical potential in the gap $%
\protect\mu=0.0$ and (b) in the valence band $\protect\mu=-2.0$. Here $mB=-2.0$ and $-0.2$
represent the topological nontrivial case and $mB=0.2$, $0.5$, and $1.0$ correspond
the topological trivial case.}
\label{fig-6}
\end{figure}

The mean field $b_{0}^{2}$ introduced in Sec. \ref{sec:sbmf} gives
the probability that the impurity is empty. Figure \ref{fig-6}
presents the dependence of $b_{0}^{2}$ on the impurity energy level
$\epsilon _{d}$ for different values of $mB$. $b_{0}^{2}=0$ when the
impurity level $\epsilon _{d}$ is much lower than the chemical
potential, which means that the impurity is singly occupied. In this
case, the charge fluctuation between the impurity and the bulk bands
is suppressed. When $\epsilon _{d}$ exceeds a threshold value,
$b_{0}^{2}$ begins to increase from 0 and saturates at 1
when $\epsilon _{d}\gg \mu $. The mixed-valance regime is defined as where $%
b_{0}^{2}$ changes from 0 to 1. Figure \ref{fig-6} presents $b^{2}$
when the chemical potential lies in the gap and in the valence
bands. As shown in Fig. \ref{fig-6}, the mixed-valence
regime is broader for the band-inverted case with $mB>0$. For the
normal case with $mB<0$, $b_{0}^{2}$ increases more rapidly, and the
mixed-valence regime shrinks into a very narrow regime. This kind of
narrow mixed-valence regime has also been found
in unconventional density waves and $d$-wave superconductors.\cite%
{Balatsky06rmp,Dora05prb} For $mB<0$, the density of states of bulk
electrons vanishes at the band edges. The reduction of the density of states
at the band edges implies that the Kondo resonance peak is narrower and $%
b_{0}^{2}$ decreases to zero faster than those in the conventional case.
From the self-consistent equations, near the critical point, the critical
value of $\epsilon _{d}$ is determined by
\begin{equation*}
\epsilon _{dc}=2\int_{-D}^{\mu }d\omega \frac{g(\omega )}{\omega -\tilde{%
\epsilon}_{d}},
\end{equation*}%
where $g(\omega )$ is proportional to the density of states of the
background electrons. In the band-inverted case, $g(\omega )$ diverges near
the band edges, which results in a large integral value as well as large $%
|\epsilon _{dc}|$. In contrast, $g(\omega )$ reduces to zero near the band
edges in the normal case. In this case, $|\epsilon _{dc}|$ is smaller and
the mixed-valence regime becomes narrower.

\subsection{In two dimensions}

The preceding sections present the numerical results in three dimensions. In
this section, we briefly discuss the behaviors of the quantum impurity in
the bulk of two-dimensional TIs. In practice, the two-dimensional case is
more accessible by the scanning tunneling miscroscope.\cite{Alpichshev12prl}
At the mean-field level, the difference between  two and three dimensions
arises from the self-energy $\Sigma _{0}(\omega )$, which in two dimensions
can be expressed as
\begin{equation*}
\Sigma _{0}(\omega )=\frac{2\omega \tilde{V}_{0}^{2}N_{0}}{k_{F}^{d}}%
\int_{0}^{\infty }dk\frac{k}{\omega ^{2}-\hbar ^{2}v_{F}^{2}k^{2}-A_{k}^{2}}.
\end{equation*}%
A detailed discussion about the properties of $\Sigma _{0}(\omega )$ in two
dimensions is presented in Ref. \onlinecite{Chan12prb}, in which a $\delta $%
-impurity scattering is considered. It can be deduced that, at the point $%
\omega =\Delta _{N}/2$, the self-energy $\Sigma _{0}(\omega )$ is finite in
three dimensions, while in two dimensions, $\Sigma _{0}(\omega )$ has
logarithmic divergence to $-(+)\infty $ at $\omega \rightarrow +(-)|m|$ for
TIs and has logarithmic divergence to $+(-)\infty $ at $\omega \rightarrow
-(+)|m|$ for normal insulators.\cite{Chan12prb} This means that, when the Anderson
impurity couples only to one band, the topological phase transition can be
seen from the position of impurity bound states as the system changes from
normal insulator to TI. Besides, the physical properties of the quantum
impurity are similar qualitatively in two and three dimensions, e.g., the
mixed-valence regime in both cases is much broader for $mB>0$ than the case
of $mB<0$.

\section{Self-screening of the Kondo effect}

\begin{figure}[htbp]
\centering \includegraphics[width=0.5\textwidth]{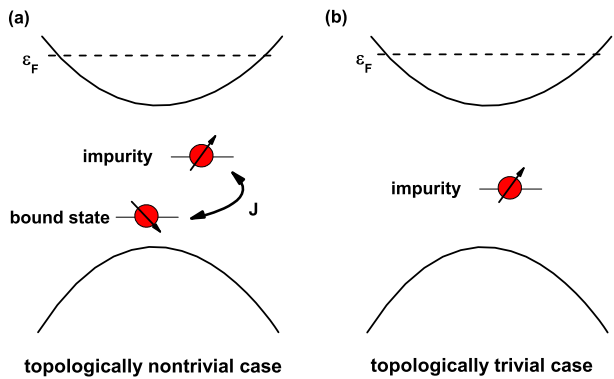}
\caption{ Illustration of the Kondo effect for (a) topological nontrivial
case and (b) trivial case.}
\label{fig:exchange}
\end{figure}

We have shown that the quantum impurity may induce in-gap bound
states in the band-inverted case, leading to the coexistence of the
Kondo resonance and the bound states. This indicates that the Kondo
resonance and bound state originate from two different mechanisms.
The singly-occupied quantum impurity behaves like a single spin, and
gives the Kondo resonance when the system is in the Kondo regime.
The bound state, on the other hand, is induced when the system is
topologically nontrivial. Even a potential scattering\cite{Liu09prb}
or a vacancy\cite{Shan11prb} could produce the in-gap bound state in
the bulk of TIs. Therefore, the in-gap bound state could be
considered as a degree of freedom, while its energy is determined by
the impurity.  When the chemical potential of the TI lies in the
conduction bands, both the bound states and the impurity level are
occupied, and each of them behaves like a single localized spin
(as illustrated in Fig. \ref{fig:exchange}). It is quite naturally
to expect the exchange interaction between two quantum spins, which
is usually due to second-order virtual hopping or the  
Ruderman-Kittel-Kasuya-Yosida (RKKY)
mechanism mediated by itinerant electrons. If the interaction is
antiferromagnetic, the two spins form a singlet, which will compete
with the many-body singlet formed by the impurity spin and
conduction electrons, quenching their Kondo effect. Different from
the intensively studied two-impurity Kondo
problem,\cite{Jones88prl,Lopez02prl} here the quenched Kondo effect
is induced by the spin of the bound states rendered by the impurity
itself. Therefore, we refer to this effect as the self-screening of
the Kondo effect.

\subsection{Exchange interaction between impurity spin and impurity-induced
bound-state spin}

The self-screened Kondo effect can be illustrated by the model Hamiltonian
as follows
\begin{eqnarray}
H&=&H_0+\sum_\sigma\epsilon_dd_{\sigma}^{\dagger}d_{\sigma}
+\sum_\sigma\epsilon_ff_{\sigma}^{\dagger}f_{\sigma}+J\mathbf{S}_d\cdot
\mathbf{S}_f  \notag \\
&&+\sum_{k\sigma}(V_{d}b_da_{k\sigma}^{\dagger}d_{\sigma}
+V_{d}b_db_{k\sigma}^{\dagger}d_{\sigma}+\mathrm{H.c.})  \notag \\
&&+\sum_{k\sigma}(V_{f}b_fa_{k\sigma}^{\dagger}f_{\sigma}
+V_{f}b_fb_{k\sigma}^{\dagger}f_{\sigma}+\mathrm{H.c.})  \notag \\
&&+\lambda_d(\sum_\sigma d_{\sigma}^{\dagger}d_{\sigma}+b_d^2-1)
+\lambda_f(\sum_\sigma f_{\sigma}^{\dagger}f_{\sigma}+b_f^2-1),  \notag \\
\end{eqnarray}
where $\epsilon_f$ is the impurity-induced bound-state level and
$H_0$ is the Hamiltonian for the bulk of the TI [Eq. (2)]. The term
$J\mathbf{S}_d\cdot \mathbf{S}_f$ describes the exchange interaction
between the impurity spin and bound-state spin. It is noted that
here the quantum impurity and its induced in-gap bound state are
considered as two degrees of freedom.

The above effective Hamiltonian describes the low-energy behavior of
the Anderson impurity in the nontrivial TI phase of the system. It
can be derived from a renormalization-group approach where the 
high-energy degree of the system is integrated out systematically. The
derivation of the effective coupling between the bound state and
Kondo resonance requires a technique beyond the slave-boson mean-field
theory, which was presented recently.\cite{Kuzmenko13etal} Using a weak
coupling renormalization group analysis, it has been shown that
the exchange interaction $J$ between the d and induced f spins
may be renormalized dynamically to either positive or negative values.
In the regime where charge fluctuations in both d and f states are
quenched, the system is in the self-screened Kondo
regime for $J>0$ and in the SO(3) Kondo regime for $J<0$,\cite{Kuzmenko13etal} respectively.
Here we introduce the exchange interaction $J$ and perform the
slave-boson approach  to describe the Kondo physics for small charge fluctuations in the regime
of $J>0$.

\subsection{Order parameter for the exchange interaction}

Similar to the treatment in Sec. \ref{sec:sbmf}, two slave-boson operators $%
b_{d}$ and $b_{f}$ are introduced to replace $c_{d(f)\sigma }$ by $%
b_{d(f)}^{\dagger }d(f)_{\sigma }$ in the large-U limit. The spin exchange
term $J\mathbf{S}_{d}\cdot \mathbf{S}_{f}=J\sum_{\sigma ,\sigma ^{\prime
}}d_{\sigma }^{\dagger }d_{\sigma ^{\prime }}f_{\sigma ^{\prime }}^{\dagger
}f_{\sigma }$ can be decoupled by introducing a valence-bond field $\Delta
_{0}=-\sum_{\sigma }\langle d_{\sigma }^{\dagger }f_{\sigma }\rangle $. In
the mean-field approximation\cite{Lopez02prl}
\begin{equation*}
J\mathbf{S}_{d}\cdot \mathbf{S}_{f}\rightarrow J\Delta _{0}\sum_{\sigma
}(d_{\sigma }^{\dagger }f_{\sigma }+f_{\sigma }^{\dagger }d_{\sigma
})+J\Delta _{0}^{2},
\end{equation*}%
the Hamiltonian becomes quadratic in the fermion operators. The problem is
still far from trivial as $b_{d}$, $b_{f}$, $\lambda _{d}$, $\lambda _{f}$, $%
\Delta _{0}$, and $\epsilon _{f}$ need to be determined self-consistently.
Different from the ordinary Kondo problem, here the bound-state energy $%
\epsilon _{f}$ also enters the self-consistent equations. By minimizing the
ground-state energy, one obtains a set of nonlinear self-consistent
equations
\begin{eqnarray}
\sum_{\sigma }\langle d_{\sigma }^{\dagger }d_{\sigma }\rangle +b_{d}^{2}-1
&=&0,  \notag \\
\sum_{\sigma }\langle f_{\sigma }^{\dagger }f_{\sigma }\rangle +b_{f}^{2}-1
&=&0,  \notag \\
\sum_{\sigma }(\langle d_{\sigma }^{\dagger }f_{\sigma }\rangle +\langle
f_{\sigma }^{\dagger }d_{\sigma }\rangle )+2\Delta _{0} &=&0,  \notag \\
\sum_{k,\sigma }(V_{d}\langle a_{k\sigma }^{\dagger }d_{\sigma }\rangle
+V_{d}\langle b_{k\sigma }^{\dagger }d_{\sigma }\rangle +h.c.)+2b_{d}\lambda
_{d} &=&0,  \notag \\
\sum_{k,\sigma }(V_{f}\langle a_{k\sigma }^{\dagger }f_{\sigma }\rangle
+V_{f}\langle b_{k\sigma }^{\dagger }f_{\sigma }\rangle +h.c.)+2b_{f}\lambda
_{f} &=&0.
\end{eqnarray}%
To simplify the calculation, we assume that the bound states are always
singly occupied ($b_{f}=0$) and decoupled from the conduction electrons. In
this case, the in-gap bound states act like a single spin. Correspondingly,
the constraints for the bound states becomes $\lambda _{f}(\sum_{\sigma
}\langle f_{\sigma }^{\dagger }f_{\sigma }\rangle -1)$ and $\lambda _{f}$ is
contained in the related Green's functions.

\subsection{Green's functions}

Performing the equation of motion procedure, we can obtain the Green's
functions for the impurity and the bound states,
\begin{eqnarray}
\langle\langle d_{\sigma}|d_\sigma^{\dagger }\rangle\rangle= \frac{1}{\omega-%
\tilde{\epsilon}_d-\frac{J^2\Delta_0^2} {\omega-\tilde{\epsilon}_f}%
-\Sigma_0(\omega)},  \notag \\
\langle\langle f_{\sigma}|f_\sigma^{\dagger }\rangle\rangle= \frac{1}{\omega-%
\tilde{\epsilon}_f-\frac{J^2\Delta_0^2} {\omega-\tilde{\epsilon}%
_d-\Sigma_0(\omega)}},
\end{eqnarray}%
where $\tilde{\epsilon}_d=\epsilon_d+\lambda_d$, $\tilde{\epsilon}%
_f=\epsilon_f+\lambda_f$, and $\Sigma_0(\omega)=\omega\sum_k\frac{2\tilde{V}%
_0^2} {\omega^2-\hbar^2v_F^2k^2-A_k^2}$.

In the limit $\Delta _{0}\rightarrow 0$, the impurity and the bound states
are decoupled and the results reduce to those when $J=0$. When the spin
exchange interaction exceeds a critical $J_{c}$, a nonzero order parameter $%
\Delta _{0}$ appears and the Kondo peak near the chemical potential is
expected to split. From the Green's functions $\langle \langle d_{\sigma
}|d_{\sigma }^{\dagger }\rangle \rangle $, one obtains the self-consistent
equation for the bound states,
\begin{equation*}
\epsilon _{f}-\tilde{\epsilon}_{d}-\frac{J^{2}\Delta _{0}^{2}}{\epsilon _{f}-%
\tilde{\epsilon}_{f}}-\Sigma _{0}(\epsilon _{f})=0.
\end{equation*}%
Defining $\alpha (\omega )=[\omega -\tilde{\epsilon}_{d}-\mathrm{Re}\Sigma
_{0}(\omega )](\omega -\tilde{\epsilon}_{f})-J^{2}\Delta _{0}^{2}$ and $%
\beta (\omega )=\mathrm{Im}\Sigma _{0}(\omega )(\omega -\tilde{\epsilon}_{f})
$, the self-consistent equations are derived as
\begin{eqnarray}
-\frac{2}{\pi }\int_{-\infty }^{\mu }d\omega \frac{(\omega -\tilde{\epsilon}%
_{f})\beta (\omega )}{\alpha (\omega )^{2}+\beta (\omega )^{2}}+b_{d}^{2}-1
&=&0,  \notag \\
-\frac{2}{\pi }\int_{-\infty }^{\mu }d\omega \frac{J^{2}\Delta _{0}^{2}%
\mathrm{Im}\Sigma _{0}(\omega )}{\alpha (\omega )^{2}+\beta (\omega )^{2}}
&=&1,  \notag \\
-\frac{2}{\pi }\int_{-\infty }^{\mu }d\omega \frac{J\Delta _{0}\beta (\omega
)}{\alpha (\omega )^{2}+\beta (\omega )^{2}}+\Delta _{0} &=&0,  \notag \\
\frac{2}{\pi }\int_{-\infty }^{\mu }d\omega \frac{\lbrack (\omega -\tilde{%
\epsilon}_{d})(\omega -\tilde{\epsilon}_{f})-J^{2}\Delta _{0}^{2}]\beta
(\omega )}{\alpha (\omega )^{2}+\beta (\omega )^{2}} &=&\lambda
_{d}b_{d}^{2},  \notag \\
\alpha (\epsilon _{f}) &=&0.
\end{eqnarray}%
This set of equations can be solved numerically.

\begin{figure}[htbp]
\centering \includegraphics[width=0.5\textwidth]{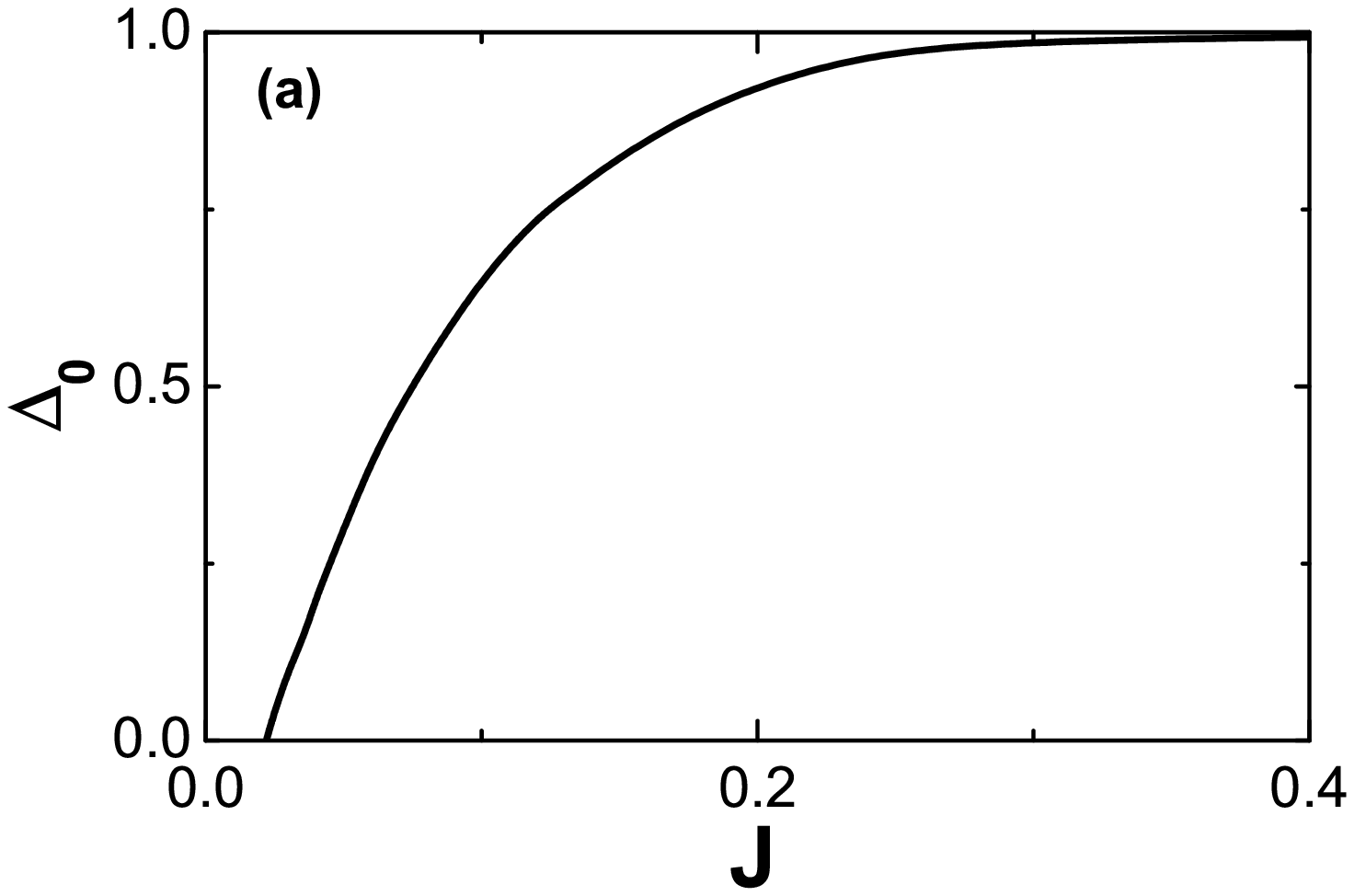} %
\includegraphics[width=0.5\textwidth]{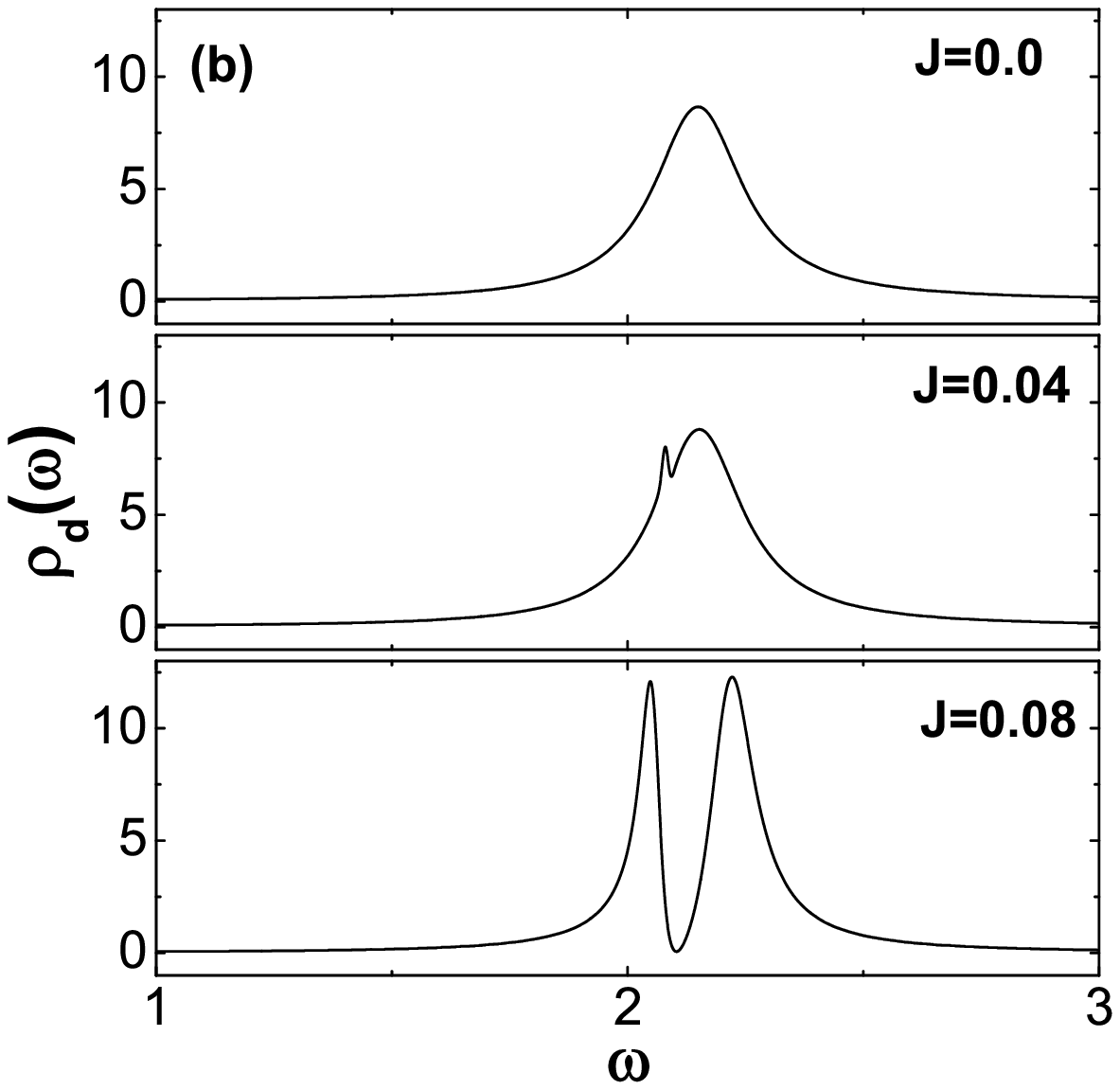}
\caption{(a) The order parameter $\Delta_0$ as a function of $J$.
Parameters: $\protect\epsilon_d=1.0$, $\protect\mu=2.0$, and $mB=1.0$. (b)
The density of states of the impurity for $J=0.0$, $0.04$, and $0.08$.}
\label{fig-8}
\end{figure}

\subsection{Self-screened Kondo effect}

Figure 8 presents the effects of bound-state spin on the Kondo effect
for different exchange interaction strength $J$. The chemical
potential of the bulk of the TI is tuned into the conduction bands, so
there is a Kondo peak near the chemical potential and the bound
states are occupied by a single
electron. When $J$ exceeds a critical value $J_{c}$, the order parameter $%
\Delta _{0}$ begins to increase from zero, then quickly to $1$ with
the increase of $J$. Figure 8(b) shows the density of states of the
impurity as a function of energy for $J=0.0$, $0.04$, and $0.08$.
With the increase of $J$, the resonance peak splits. The splitting
of the Kondo peak increases with $J$. As a result, the density of states
near $\tilde{\epsilon}_{d}$ reduces to a very small value,
corresponding to the self-screening of the Kondo resonance.

It should be noted that the strength of exchange interaction between
the impurity and the in-gap bound state is the key parameter of the
predicted self-screened Kondo effect. The exchange strength should
be evaluated subtly, for instance, from the first-principles 
calculation. Our calculation indicates that a small exchange
interaction could make the Kondo effect break down.

Above we assumed that the bound states are singly-occupied in the large-U
limit. If the Coulomb repulsion energy on the bound states is finite and the
chemical potential is high enough, it is possible that the bound states are
occupied by two electrons, and they form a singlet due to the Pauli
exclusion principle. In this limit, the bound-state spin is decoupled from
the impurity spin, and the Kondo effect originating from the interaction
between the impurity spin and conduction electrons can recover.

An impurity- or vacancy- induced in-gap bound state is a special feature
of TIs, which is absent in  normal insulators. Therefore, the
bound state may play an important role when the Kondo
physics is considered in TIs. For a strong exchange interaction between impurity and
in-gap bound state, the Kondo effect may be broken down when the
chemical potential is tuned properly. Actually, several experiments
have been performed to investigate the physical properties of
magnetic impurity-doped topological materials, such as Mn-doped
BiTe.\cite{Hor10prb} The self-screening effect is expected to be
observed in these systems.

\section{Summary}

In summary, we have studied the Kondo effect and the formation of
in-gap bound states induced by an Anderson impurity coupled with the
bulk states of topological insulators. It is demonstrated that the
positions of the Kondo peak and bound states strongly depend on the
topological properties, the chemical potential, and other parameters
of the system. The behaviors of the resonance level in the bulk of
TIs differ from those for simple metals and normal insulators. Due
to the divergence of the density of states near the band edges, the
mixed-valence regime is much broader in the band-inverted case,
while it shrinks to a very narrow regime in the normal case. For the
band-inverted case, the in-gap bound states and the Kondo resonance
can coexist. However, only one of them exists in the normal
insulators. When the impurity energy level is far away from the
chemical potential, the in-gap bound states are very close to the
band edges and can be considered as merging into the bulk.
Furthermore, we show that a self-screening Kondo effect may be
induced by taking the interaction between the impurity spin and
bound-state spin into account.

Note added: while this manuscript was under review, we became aware of
a work [50], wherein the scattering of dilute magnetic
impurities placed on the surface of TIs is investigated.

This project was supported by the Research Grant Council under Grants No. HKU
7051/10P and No. HKUST3/CRF/09, the Foundation for Innovative Research Groups of
the NSFC under Grant No. 61021061, the NSFC under Grants No. 11004022 and No. 61006081.

\end{document}